\documentclass[amsmath,amssymb,aps,prd,10pt]{revtex4-2}

\usepackage[T1]{fontenc}
\usepackage{color}
\usepackage{graphicx}
\usepackage{mathtools}
\usepackage{ragged2e}

\usepackage{dcolumn}
\usepackage{bm}
\usepackage{graphicx,empheq}
\usepackage{verbatim} 
\usepackage[english]{babel}
\usepackage{hyperref}

\hypersetup{
citecolor=red,
colorlinks=true,
filecolor=red,
linkcolor=blue,
linktocpage=true,
urlcolor=blue
} 
\usepackage[titletoc,toc,title]{appendix}
\numberwithin{equation}{section}
\usepackage{cleveref}
\usepackage{epsfig}
\usepackage{amsfonts}
\usepackage{enumitem}

\newcommand\trick[1]{}
\newcommand{\be}{\begin{equation}} 
\newcommand{\ee}{\end{equation}}
\newcommand{\eq}[1]{(\ref{#1})}
\newcommand{\bit}{\begin{itemize}}  \newcommand{\eit}{\end{itemize}}
\newcommand{\ben}{\begin{enumerate}}  \newcommand{\een}{\end{enumerate}}

\newcommand{\ket}[1]{|#1 \rangle}

\newcommand{\rf}[1]{(\ref{#1})}

\def\bd{\begin{document}}
\def\ed{\end{document}}
\def\bea{\begin{eqnarray}}
\def\eea{\end{eqnarray}}
\let\bm=\bibitem

\def\la{\langle}
\def\ra{\rangle}

\def\npb#1#2#3{Nucl. Phys. {\bf{B#1}} #3 (#2)}
\def\plb#1#2#3{Phys. Lett. {\bf{#1B}} #3 (#2)}
\def\prl#1#2#3{Phys. Rev. Lett. {\bf{#1}} #3 (#2)}
\def\prd#1#2#3{Phys. Rev. {D bf{#1}} #3 (#2)}
\def\cmp#1#2#3{Comm. Math. Phys. {\bf{#1}} #3 (#2)}
\def\cqg#1#2#3{Class. Quantum Grav. {\bf{#1}} #3 (#2)}
\def\nppsa#1#2#3{Nucl. Phys. B (Proc. Suppl.) {\bf{#1A}}#3 (#2)}
\def\ap#1#2#3{Ann. of Phys. {\bf{#1}} #3 (#2)}
\def\ijmp#1#2#3{Int. J. Mod. Phys. {\bf{A#1}} #3 (#2)}
\def\rmp#1#2#3{Rev. Mod. Phys. {\bf{#1}} #3 (#2)}
\def\mpla#1#2#3{Mod. Phys. Lett. {\bf A#1} #3 (#2)}
\def\jhep#1#2#3{J. High Energy Phys. {\bf #1} #3 (#2)}
\def\atmp#1#2#3{Adv. Theor. Math. Phys. {\bf #1} #3 (#2)}

\def\sst{\scriptscriptstyle}
\def\thetabar{\bar\theta}
\def\Tr{{\rm Tr}}
\def\one{\mbox{1 \kern-.59em {\rm l}}}

\def\a{\alpha}      \def\da{{\dot\alpha}}  \def\dA{{\dot A}}
\def\b{\beta}       \def\db{{\dot\beta}}
\def\g{\gamma}  \def\G{\Gamma}  \def\dc{{\dot\gamma}}
\def\d{\delta}  \def\D{\Delta}  \def\ddt{\dot\delta}
\def\e{\epsilon}
\def\ve{\varepsilon}
\def\uve{\upvarepsilon}
\def\f{\phi}    \def\F{\Phi}    \def\vvf{\f}
\def\vphi{\varphi}
\def\h{\eta}
\def\k{\kappa}
\def\l{\lambda} \def\L{\Lambda}
\def\m{\mu} \def\n{\nu}
\def\o{\omega}
\def\p{\pi} \def\P{\Pi}
\def\r{\rho}
\def\s{\sigma}  \def\S{\Sigma}
\def\t{\tau}
\def\th{\theta} \def\Th{\Theta} \def\vth{\vartheta}
\def\X{\Xeta}
\def\z{\zeta}

\def\na{\nabla}

\def\cA{{\cal A}} \def\cB{{\cal B}} \def\cC{{\cal C}}
\def\cD{{\cal D}} \def\cE{{\cal E}} \def\cF{{\cal F}}
\def\cG{{\cal G}} \def\cH{{\cal H}} \def\cI{{\cal I}}
\def\cJ{{\mathscr J}} \def\cK{{\cal K}} \def\cL{{\cal L}}
\def\cM{{\cal M}} \def\cN{{\cal N}} \def\cO{{\cal O}}
\def\cP{{\cal P}} \def\cQ{{\cal Q}} \def\cR{{\cal R}}
\def\cS{{\cal S}} \def\cT{{\cal T}} \def\cU{{\cal U}}
\def\cV{{\cal V}} \def\cW{{\cal W}} \def\cX{{\cal X}}
\def\cY{{\cal Y}} \def\cZ{{\cal Z}}
\def\ct{{\cal t}}

\def\ua{\underline{\alpha}}
\def\uc{\underline{\phantom{\alpha}}\!\!\!\gamma}
\def\um{\underline{\mu}}
\def\ud{\underline\delta}
\def\ue{\underline\epsilon}
\def\una{\underline a}\def\unA{\underline A}
\def\unb{\underline b}\def\unB{\underline B}
\def\unc{\underline c}\def\unC{\underline C}
\def\und{\underline d}\def\unD{\underline D}
\def\une{\underline e}\def\unE{\underline E}
\def\unf{\underline{\phantom{e}}\!\!\!\! f}\def\unF{\underline F}
\def\unm{\underline m}\def\unM{{\underline M}}
\def\unn{\underline n}\def\unN{{\underline N}}
\def\unp{\underline{\phantom{a}}\!\!\! p}\def\unP{\underline P}
\def\unq{\underline{\phantom{a}}\!\!\! q}
\def\unQ{\underline{\phantom{A}}\!\!\!\! Q}
\def\unH{\underline{H}}

\def\As {{A \hspace{-6.4pt} \slash}\;}
\def\bs {{b \hspace{-6.4pt} \slash}\;}
\def\Ds {{D \hspace{-6.4pt} \slash}\;}
\def\Gts {{\Gt \hspace{-6.4pt} \slash}\;}
\def\ds {{\del \hspace{-6.4pt} \slash}\;}
\def\ss {{\s \hspace{-6.4pt} \slash}\;}
\def\ks {{ k \hspace{-6.4pt} \slash}\;}
\def\ps {{p \hspace{-6.4pt} \slash}\;}
\def\xs {{x \hspace{-6.4pt} \slash}\;}
\def\pas {{{p_1} \hspace{-6.4pt} \slash}\;}
\def\pbs {{{p_2} \hspace{-6.4pt} \slash}\;}
\def\cFs {{{\cal F} \hspace{-6.4pt} \slash}\;}
\def\Dss {{D \hspace{-7.5pt} \slash}\;}
\def\dss {{\del \hspace{-7.0pt} \slash}\;}

\def\Ah{{\hat{A}}}
\def\Dh{{\hat{D}}}
\def\Gh{{\hat{G}}}
\def\Fh{{\hat{F}}}
\def\Ih{{\hat{I}}}
\def\Jh{{\hat{J}}}
\def\Kh{{\hat{K}}}
\def\Lh{{\hat{L}}}
\def\Ph{{\hat{P}}}
\def\Rh{{\hat{R}}}
\def\Vh{{\hat{V}}}
\def\Xh{{\hat{X}}}

\def\ah{{\hat{\a}}}
\def\bh{{\hat{\b}}}
\def\gh{{\hat{\g}}}
\def\dh{{\hat{\d}}}
\def\rh{{\hat{\r}}}
\def\hh{\hat{h}}
\def\uh{\hat{u}}
\def\xh{\hat{x}}
\def\yh{\hat{y}}
\def\ph{\hat{p}}
\def\xih{\hat{\xi}}
\def\chih{\hat{\chi}}
\def\Psih{\hat{\Psi}}
\def\phih{\hat{\phi}}

\def\psit{\tilde{\psi}}
\def\Psit{\tilde{\Psi}}
\def\Psibt{\tilde{\bar{Psi}}}

\def\lambdat{\tilde {\lambda}}
\def\st{\tilde{\sigma}}

\def\delt{\tilde{\delta}}
\def\Phit{\tilde{\Phi}}
\def\Phitb{\overline{\tilde{Phi}}}
\def\tht{\tilde{\th}}
\def\lt{\tilde{\l}}
\def\chit{\tilde{\chi}}
\def\phit{\tilde{\phi}}

\def\At{\tilde{A}}
\def\Bt{\tilde{B}}
\def\Ct{\tilde{C}}
\def\Dt{\tilde{D}}
\def\Et{\tilde{E}}
\def\Ft{\tilde{F}}
\def\Gt{\tilde{G}}
\def\Ht{\tilde{H}}
\def\It{\tilde{I}}
\def\Jt{\tilde{J}}
\def\Qt{\tilde{Q}}
\def\Rt{\tilde{R}}
\def\Mt{\tilde{M }}
\def\Nt{\tilde{N}}
\def\St{\tilde{S}}
\def\Vt{\tilde{V}}
\def\Xt{\tilde{X}}
\def\at{\tilde{a}}
\def\dt{\tilde{d}}
\def\htt{\tilde{h}}
\def\ft{\tilde{f}}
\def\gt{\tilde{g}}
\def\pt{\tilde{p}}
\def\qt{\tilde{q}}
\def\vt{\tilde{v}}
\def\nt{\tilde{n}}
\def\ut{\tilde{u}}
\def\wt{\tilde{w}}
\def\zt{\tilde{z}}
\def\xt{\tilde{x}}
\def\yt{\tilde{y}}
\def\Psit{\tilde{\Psi}}
\def\vphit{\tilde{\varphi}}
\def\tD{\tilde{\D}}

\def\eb{\bar{\epsilon}}
\def\delb{\bar{\partial}}
\def\thb{\bar{\theta}}
\def\mub{\bar{\mu}}
\def\lamb{\bar{\l}}
\def\psib{\bar{\psi}}
\def\sb{\bar{\sigma}}
\def\xib{\bar{\xi}}
\def\chib{\bar{\chi}}

\def\Psib{\bar{\Psi}}
\def\Phib{\bar{\Phi}}
\def\Lamb{\bar{\Lambda}}
\def\Sb{{\overline \Sigma}}
\def\cb{\bar{c}}
\def\hb{\bar{h}}
\def\qb{\bar{q}}
\def\wb{\bar{w}}
\def\ub{\bar{u}}
\def\zb{{\bar{z}}}
\def\Hb{\bar{H}}
\def\Qb{{\bar Q}}
\def\Omegab{\overline{\Omega}}
\def\ob{\overline{\omega}}

\def\Ab{{\overline A}} \def\Bb{{\overline B}} \def\Cb{{\overline C}}
\def\Db{{\overline D}} \def\Eb{{\overline E}} \def\Fb{{\overline F}}
\def\Gb{{\overline G}}
\def\Ib{{\overline I}}
\def\Jb{{\overline J}} \def\Kb{{\overline K}} \def\Lb{{\overline L}}
\def\Mb{{\overline M}} \def\Nb{{\overline N}} \def\Ob{{\overline O}}
\def\Pb{{\overline P}}  \def\Rb{{\overline R}}
 \def\Tb{{\overline T}} \def\Ub{{\overline U}}
\def\Vb{{\overline V}} \def\Wb{{\overline W}} \def\Xb{{\overline X}}
\def\Yb{{\overline Y}} \def\Zb{{\overline Z}}

\def\fb{{\overline f}}
\def\gb{{\overline g}}
\def\nb{{\overline n}}
\def\mb{{\overline m}}
\def\lb{{\overline l}}
\def\yb{{\overline y}}

\def\ldel{{\overleftarrow{\del}}}
\def\rdel{{\overrightarrow{\del}}}
\def\ldeldel{{\overleftarrow{\del^2}}}
\def\rdeldel{{\overrightarrow{\del^2}}}
\def\ldelb{{\overleftarrow{\bar{\del}}}}
\def\rdelb{{\overrightarrow{\bar{\del}}}}

\def\ba{{\bf a}}
\def\bk{{\bf k}}
\def\bl{{\bf l}}
\def\bp{{\bf p}}
\def\bq{{\bf q}}
\def\br{{\bf r}}
\def\bt{{\bf t}}
\def\bu{{\bf u}}
\def\bv{{\bf v}}
\def\bx{{\bf x}}
\def\by{{\bf y}}
\def\bA{{\bf A}}
\def\bR{{\bf R}}
\def\bV{{\bf V}}

\def\bz{{\boldsymbol{\zeta}}}

\def\bone{{\bf 1}}

\def\va{{\vec a}}
\def\vk{{\vec k}}
\def\vp{{\vec p}}
\def\vq{{\vec q}}
\def\vx{{\vec x}}
\def\vy{{\vec y}}
\def\vu{{\vec u}}
\def\vv{{\vec v}}
\def \vH{{\vec H}}
\def \vg{{\vec g}}

\def\vs{{\vec \sigma}}
\def\vtau{{\vec \tau}}

\newcommand{\ov}[1]{\overrightarrow{#1}}

\def\frA{\mathfrak{A}}
\def\frB{\mathfrak{B}}
\def\frC{\mathfrak{C}}
\def\frD{\mathfrak{D}}
\def\frE{\mathfrak{E}}
\def\frF{\mathfrak{F}}
\def\frG{\mathfrak{G}}
\def\frH{\mathfrak{H}}
\def\frM{\mathfrak{M}}
\def\frN{\mathfrak{N}}
\def\frR{\mathfrak{R}}
\def\frW{\mathfrak{W}}

\def\fra{\mathfrak{a}}
\def\frb{\mathfrak{b}}
\def\frf{\mathfrak{f}}
\def\frg{\mathfrak{g}}
\def\frh{\mathfrak{h}}
\def\frl{\mathfrak{l}}
\def\frs{\mathfrak{s}}
\def\fri{\mathfrak{i}}
\def\frj{\mathfrak{j}}

\def\ma{\mathfrak{a}}
\def\mg{\mathfrak{g}}
\def\mh{\mathfrak{h}}
\def\mR{\mathfrak{R}}
\def\mN{\mathfrak{N}}

\newcommand{\nn}{{\nonumber}}

\def\d{\delta}\def\D{\Delta}\def\ddt{\dot\delta}

\def\pa{\partial} \def\del{\partial}
\def\xx{\times}
\def\uno{\mbox{1 \kern-.59em {\rm l}}}

\def\trp{^{\top}}
\def\inv{^{-1}}
\def\dag{\dagger}
\def\pr{^{\prime}}

\def\rar{\rightarrow}
\def\lar{\leftarrow}
\def\lrar{\leftrightarrow}

\newcommand{\0}{\,\!}      
\def\one{1\!\!1\,\,}
\def\im{\imath}
\def\jm{\jmath}

\newcommand{\tr}{\mbox{tr}}
\newcommand{\slsh}[1]{/ \!\!\!\! #1}

\newcommand{\1}{\mbox{1}\hspace{-0.25em}\mbox{l}}

\def\vac{|0\rangle}
\def\lvac{\langle 0|}

\def\hlf{\frac{1}{2}}
\def\ove#1{\frac{1}{#1}}
\newcommand{\hot}[1]{\frac{#1}{2}}

\def\Box{\square}
\def\CC {\mathbb{C}}
\def\FF {\mathbb{F}}
\def\RR{\mathbb{R}}
\def\NN{\mathbb{N}}
\def\ZZ{\mathbb{Z}}
\def\bb#1{{\bf #1}}
\def\bcomment#1{}
\def\bfhat#1{{\bf \hat{#1}}}
\def\VEV#1{\left\langle #1\right\rangle}

\newcommand{\ex}[1]{{\rm e}^{#1}} \def\ii{{\rm i}}

\newcommand{\lrbrk}[1]{\left(#1\right)}
\newcommand{\lrsbrk}[1]{\left[#1\right]}
\newcommand{\sfrac}[2]{{\textstyle\frac{#1}{#2}}}

\def\stw{{\sqrt{2}}}

\def\rf {{\rm f}}
\def\ri {{\rm i}}
\def\rj {{\rm j}}
\def\rn {{\rm n}}
\def\rk {{\rm k}}
\def\rl {{\rm l}}
\def\rr {{\rm r}}
\def\rs {{\scriptscriptstyle \rm S}}
\def\rt {{\scriptscriptstyle \rm T}}

\def\rQ {{\scriptscriptstyle \rm \cQ}}
\def\rR {{\scriptscriptstyle \rm \cR}}

\def\cQb{{\cal \Qb}}
\def\cRb{{\cal \Rb}}
\def\cWb{{\cal \Wb}}

\def\fd {{\rm N}}
\def\afd {{\overline{\rm N}}}

\def \II {I\hspace{-.1em}I\hspace{.1em}}
\def \IIA {\mbox{\II A\hspace{.2em}}}
\def \IIB {\mbox{\II B\hspace{.2em}}}
\def \gs {g^s}
\def \ls {\lambda^s}

\def \I {{\cal I}}
\def \qs {q\hspace{-.53em}/\hspace{.15em}}
\def \ks {k\hspace{-.53em}/\hspace{.15em}}
\def \YM {{\mbox{\tiny YM}}}
\def \gym {g_{\YM}}

\def \Lc {\L_c}
\def\IR{\relax{\rm I\kern-.18em R}}
\def \id {{\bf 1}}

\def\cci{\ell}
\def\ccj{\ell'}

\def\bbq{\pmb{q}}
\def\bom{\pmb{\o}}
\def\bJ{\pmb{J}}
\def\bM{\pmb{M}}
\def\bB{\pmb{B}}
\def\bn{\pmb{n}}
\def\bE{\pmb{E}}

\newcommand{\rrr}[1]{\vskip 0.2cm \noindent{\bf #1} ---}

\long\def\symbolfootnote[#1]#2{\begingroup%
\def\thefootnote{\fnsymbol{footnote}}\footnote[#1]{#2}\endgroup}
\long\def\RemarkBox#1{\begin{flushleft}\fbox{\begin{minipage}
{17.5cm}{\bf Remark:} ~#1\end{minipage}}\end{flushleft}}
\newcommand{\aei}{\it Max Planck Institute for Gravitational Physics
(Albert Einstein Institute)\\ Am M\"uhlenberg 1, 14476 Golm,
Germany}

\newcommand{\nthu}{{\it Department of Physics, National Tsing-Hua
  University,
  Hsinchu 30013, Taiwan}}

\newcommand{\ctc}{{\it
Center of Theory and Computation, 
National Tsing-Hua University, Hsinchu 30013, Taiwan}}

\newcommand{\sysu}{{\it School of Physics and Astronomy, Sun Yat-Sen
    University,
    Zhuhai 519082, China}}

\newcommand{\ncts}{{\it
National Center for Theoretical Sciences, Taipei 10617, Taiwan}}

\begin{document}
\title{A Fermi Model of Quantum Black Hole}
\author{Chong-Sun Chu}
\email[Correspondence email address: ]{cschu@phys.nthu.edu.tw}
\affiliation{ Department of Physics, National Tsing-Hua
  University,  Hsinchu 30013, Taiwan}
\affiliation{Center of Theory and Computation, 
National Tsing-Hua University, Hsinchu 30013, Taiwan}
\affiliation{National Center for Theoretical Sciences, Taipei 10617, Taiwan}

\author{Rong-Xin Miao}
\email{miaorx@mail.sysu.edu.cn} 
\affiliation{School of Physics and Astronomy, Sun Yat-Sen
    University,    Zhuhai 519082, China}

\begin{abstract}
We propose a quantum model of the Schwarzschild black hole as a
quantum mechanics of a system of fermionic degrees of freedom.  The
system has a constant density of states and a Fermi energy that is
inversely proportional to the size of the system.  Assuming
equivalence principle, we show that the degeneracy pressure of the
Fermi degrees of freedom is able to withstand the collapse of gravity
if the radius of the system is given precisely by the horizon radius
of the Schwarzschild black hole.  In our model, the fermionic degrees
of freedom at each energy level can be entangled in certain different
ways, giving rise to a multitude of degenerate ground states of the
system.  The counting of these microstates reproduces precisely the
Bekenstein-Hawking entropy.  This simple Fermi model is universal and
works also for the Reissner-Nordstr\"om charged black hole as well as
black hole with a cosmological constant.  From the properties of the
Fermi variables, we propose that quantum gravity is characterized by a
principle of {\it maximal capacity of states} where there can be no
more than $V /l_P^3$ quantum states in any volume $V$. It implies a
loss of spatial locality below the Planck length and suggests that any
singularity predicted by general relativity is resolved and replaced
by a quantum space in quantum gravity.  In our model, a black hole
spacetime is equipped with an uniform distribution of energy
levels. This is another reason why black hole can be considered a
simple harmonic oscillator of quantum gravity.

\end{abstract}

\maketitle



\section{Introduction}

The nature of quantum spacetime and quantum gravity is one of the
most important questions for fundamental physics.
Black hole arises as  solution
of the classical general relativity.  However, the quantum property of
black hole is puzzling and it is believed that only in
a full theory of quantum gravity can it be properly understood.
In this paper we propose a model
of the interior of black hole as a bottom-up approach
to learn about the construction of quantum gravity.
In our model, the following puzzling properties of
black hole are addressed and resolved.

\noindent {\it  (i) Bekenstein-Hawking Entropy:}
The laws of black hole mechanics \cite{Bardeen:1973gs}, for example, the
first law
\be \label{BHM}
dU = \frac{\k}{8 \pi G} dA + \cdots
\ee
was originally derived from the classical theory of Einstein gravity.
That these mechanical relations are, in fact, thermodynamic relations
was made evident with the discovery of the quantum
Hawking radiation \cite{Hawking:1975vcx}, which makes the
black hole to appear like a thermal object to the outside world 
with the Hawking temperature
\be \label{TH}
T_H = \frac{1}{8\pi GM}. 
\ee
As a result, the perturbation relation \eq{BHM}
becomes the 
first law of thermodynamics if the black hole carries the Bekenstein-Hawking
entropy 
\cite{Bekenstein:1972tm,Bekenstein:1973ur}
\be \label{BH}
S_{\rm BH} = \frac{A}{4G}.
\ee
The thermodynamic nature of black hole mechanics
has led Jacobson to the very interesting proposal \cite{Jacobson:1995ab}
that Einstein's theory of gravity is a
thermodynamic expression of spacetime.

The possession of the entropy \eq{BH} by a quantum black hole is
puzzling. Instead of being extensive as in standard thermodynamic
systems, the area dependence of the Bekenstein-Hawking entropy \eq{BH}
has suggested holography \cite{tHooft:1993dmi,Susskind:1994vu} as a
fundamental principle for quantum gravity and has inspired the
formulation of the AdS/CFT correspondence \cite{Maldacena:1997re}.  In
effect, any quantum gravitational theory of black holes should have
the gravitational degrees of freedom inside the black hole properly
identified such that an account of them would give rise to the
area-dependent entropy \eq{BH} of black holes. Therefore, the
reproduction of \eq{BH} becomes an important testing stone for a
consistent model of quantum black holes. See, for example,
\cite{Strominger:1996sh} for progress in microstates counting in
supersymmetric brane black hole and \cite{Mathur:2005zp} for the fuzzy
ball proposal.

\noindent {\it   (ii) Horizon Radius:}
Classically, black hole is a vacuum solution of
spacetime. 
Outside the horizon, 
black hole behaves very much like an ordinary gravitational
massive object. For the Schwarzschild black hole,
the horizon radius depends on the mass $M$ in a
specific way:
\be \label{RM}
R= 2GM := R_S,
\ee
where $R_S$ is the Schwarzschild radius.
This is strikingly different from the typical relation
$R \sim M^{-\b}, \b >0$ 
(e.g. $\b = 1/3 $ for a non-relativistic neutron star) for a
stellar objects made of ordinary matter
without any internal energy source, e.g. fusion.
That the radius is inversely correlated with the mass is  a  simple
consequence of the fact that such matter becomes
more compactly packed as the force of gravity increases. On the other hand,
the unusual behavior \eq{RM} that a black hole  gets
bigger as it get more massive  means that  the degrees of freedom of
the black hole cannot be of an ordinary variety.
For a charged black hole, the horizon radius satisfies the relation
\be \label{RMQ}
R^2 -R R_S + R_Q^2 =0,
\ee
where $R_Q := \sqrt{G Q^2}$ is the charge radius for the Reissner-Nordstr\"om
black hole.
Likewise, the horizon radius $R$ of a black hole with a cosmological
constant $\L$ satisfies the relation
\be \label{RLambda}
R -R_S - \frac{\L}{3}R^3 =0.
\ee
As a result,
the reproduction of relations \eq{RM}, \eq{RMQ} and \eq{RLambda}
for the horizon radius in the respective cases
is another important testing stone for a consistent
quantum model of  black hole.

\noindent {\it (iii)
Non-singular interior:} 
Most troubling is that as long as certain energy condition of matter
holds, the prediction of black hole singularity in general relativity
cannot be avoided
\cite{Penrose:1964wq}. Although it is generally expected that quantum
gravity will play a crucial role in resolving the singularity of
classical spacetime and provide a complete quantum description of
black holes, there are different ways it could happen. It is possible
that the Einstein equation is modified so that the curvature
singularity is avoided. A more dramatic possibility is that the
classical spacetime of the black hole is replaced by a
completely novel version of quantum existence where
the singularity theorem does not apply. This possibility is
interesting not just because it offers a valuable window to look into
the novel properties of quantum spacetime but also because, in a
quantum theory of spacetime, the black hole interior would
automatically be equipped with a Hilbert space of states, which
potentially could explain the origin of the black hole entropy
\eq{BH}.

\noindent {\it (iv) Unitarity:} Assuming quantum mechanics holds, 
black hole should be described by a unitary quantum system.
However,  apparently a black hole form in a pure
state can evolve into a mixed state as it ends its life through the
evaporation of thermal Hawking radiation \cite{Hawking:1976ra}.
This information problem contradicts the unitarity
of quantum mechanics.
Unitarity also requires the entanglement entropy of
the Hawking radiation to follow a Page curve
\cite{Page:1993wv,Page:2013dx} instead of a Hawking curve.
The recovery of quantum information and the
reproduction of the Page curve behavior of the Hawking radiation
entropy  are necessary
requirements for a consistent quantum model of black holes.

It is clear from the above discussion that the crux of all these
problems is a lack of understanding of the quantum properties of the
interior of black hole.  In principle, a consistent quantum model of
black hole without singularity is needed in order to allow for a
proper identification of the Hilbert space of states of the black
hole, and in turn this is needed to express the quantum states of the
black hole interior spacetime and which in turn is needed to explain
the Bekenstein-Hawking entropy \eq{BH}. And again, only with the
availability of the black hole Hilbert space does one has the
necessary states to possibly describe the unitarity evolution of the
black hole without the risk of missing anything.

So far the most commonly adopted approach to the problem of black hole
is a top-down one where one starts with a candidate theory of quantum
gravity and studies the properties of the black hole as a solution of
it.  For example, brane models of black hole have been considered in
non-perturbative string theory, with the Berkenstein-Hawking entropy
reproduced successfully from a microstates counting
\cite{Strominger:1996sh,Mathur:2005zp}.  Recently progress has been
achieved for black hole constructed in the AdS/CFT correspondence
\cite{Maldacena:1997re} where the Page curve behavior of the
entanglement entropy of the Hawking radiation is remarkably obtained
\cite{Engelhardt:2014gca,Penington:2019npb,Almheiri:2019psf,Almheiri:2019hni,
  Almheiri:2020cfm}.  Nevertheless, despite these success, it is still
highly desirable to have a direct formulation of the quantum gravity
itself and be able to describe the fundamental degrees of freedom of
quantized spacetime and the set of microstates directly without
resorting to supersymmetry or duality.

In a recent paper \cite{Chu:2022ieq}, we have initiated an
investigation of quantum black hole along these lines.  There, we
proposed that the interior of black hole is filled with a thin shell of
entangled Bell states located just underneath the
horizon. We argued that the configuration can be stabilized by a new
kind of degeneracy pressure originated from the noncommutative
spacetime.  We showed that quantum tunneling
\cite{Parikh:1999mf} which occurs near the horizon results in the Page
curve behavior of the entanglement entropy of the Hawking radiation,
just as islands in the AdS/CFT
\cite{Engelhardt:2014gca,Penington:2019npb,Almheiri:2019psf,
  Almheiri:2019hni,Almheiri:2020cfm}.
Moreover it was shown that the entanglement information initially
stored in the black hole can all be returned to the environment at the
end of life of the black hole.  That the origin of the Page curve and
the return of information can be elucidated in terms of explicit
spacetime quantum mechanical processes is interesting.  However, some
of the basic properties of the model in \cite{Chu:2022ieq} were
assumed and not derived. For example, how does the degeneracy pressure
arise? what kind of quanta are these?  To provide a more complete
model from which the results of \cite{Chu:2022ieq} can be derived and
firmer justified is a main motivation of this work.

Instead of a top-down approach where the fundamental theory is
assumed, we will take a bottom-up approach to use the black
hole properties as ``empirical input'' to help us 
to learn about the fundamental theory of quantum gravity.
The only assumption we make is that quantum gravity
can be formulated in terms of a quantum mechanics
of fermionic and bosonic degrees of freedom.
Generically, the Lagrangian takes the form
\be \label{matrix model}
L= i \psi^+ \dot{\psi}+  \psi^+ h(X) \psi-V(X),
\ee 
where $ h(X) $ denotes some Yukawa coupling and $V(X)$ the self-interaction.
We emphasis that these degrees of freedom are not
particles, nor are they described by an energy-momentum tensor, both
notions of which would only make sense when there is a spacetime.
Interestingly, Maldacena has
recently proposed quantum mechanics of oscillators and Majorana
fermions to describe a black hole \cite{Maldacena:2023acv}. It is
interesting that the fermionic nature of black holes is
also recognized in their considerations.

Here is the plan of the paper.
In section 2.1, we perform a general analysis for
a quantum Fermi system described 
by a density of energy eigenstates. We show that it admits
a degenerate pressure in general and there is no need to assume the degrees
of freedom are particles.
In section 2.2, we apply it to the system
of neutron and obtain the usual mass radius relation of neutron star. 
In section 2.3, we consider a quantum mechanical system of fermionic
degrees of freedom as a
model of the Schwarzschild
black hole. Outside the system, we assume that  the
low energy effective description emerges
where the usual spacetime and general relativity become valid and 
the system is seen by the outside observer as an object with a
gravitational mass $M$ and some horizon radius $R$.
We show that the desired properties of the quantum black hole
can be precisely reproduced by the microscopic Fermi system.
In fact, by assuming the equivalence principle, we show that the
degeneracy pressure of the Fermi degrees of freedom
is just right to withstand the pressure of gravitational collapse
when the size of the Fermi system is precisely given by
the Schwarzschild horizon radius \eq{RM}.
We also show that the Fermi system admits
a large number of degenerate ground states, each of which is given by a
richly entangled configuration of the bosonic and fermionic degrees of freedom;
and the counting of the ground state degeneracy reproduces
precisely the Bekenstein-Hawking entropy \eq{BH}.
This gives evidence that our
Fermi model has indeed captured the microstates of the black
hole. 
The Page curve behavior of the Hawking entropy is also reproduced in our
model \cite{Chu:2023agv}.
In section 2.4 and 2.5, we extend the analysis to 
the more general cases of a black hole with
charge or with a cosmological constant \footnote{Our model works in
principle also for rotational black hole. However, the analysis
becomes a little complicated that we do not carry it out here.}
and show that all the desired properties of the black hole are reproduced.
That the properties of various black holes
could be accounted for universally provides support to the
validity of our description.
Our model
suggests that holography is a consequence
of the fermionic nature of the quantum spacetime.
Our analysis also suggests that  quantum
gravity is characterized by a  principle of {\it maximal capacity of states}
where there can be no more than $V/l_p^3$ of
quantum states in any volume $V$, see \eq{Nmax}. This is satisfied by
our model of black hole.  The loss of
locality below the Planck length suggests that
the usual singularity theorem is avoided and our model is consistent.
Further discussions are presented in section
3.

\section{A Fermionic Model for the Quantum Black Hole}

Black hole in general relativity is described by a metric with a
singularity. The existence of singularity means that the classical
model of black hole as a vacuum spacetime is not trustable. The
formation of singularity may be avoided if the gravitational collapse
is somehow counter balanced by some internal pressure. Classically
this is impossible due to the singularity theorem \cite{Penrose:1964wq}.
In the following we consider a replacement of the
interior spacetime of black hole by a quantum mechanical
system of fermionic degrees of freedom. To an outside observer, the
system occupies a volume of space and possess an internal energy. 
We will show that the volumetric response of the
internal energy  creates a pressure
that is just right to precisely balance out the pressure 
from the gravitational spacetime outside.
We emphasis that we are not considering some field
theoretic exotic matter or vacuum system of the black hole,
see e.g. \cite{Balasin:1993fn,Roupas:2022gee,Volovik:2023phl}. Instead
we propose that the classical interior
spacetime of black hole is replaced by a quantum space and the fermionic
degrees of freedom describe the quantum fluctuations over it.

\subsection{Degenerate Fermi system}

To start with, let us determine the degeneracy
pressure of a system of Fermi degrees of
freedom. 
Consider a system of fermionic
degrees of freedom contained in a spatial volume $V$.
An important specification of
the system is the density of energy eigenstates
$g(E)$ where $g(E) dE$ is the number of
energy eigenstates contained within the energy interval $(E, E+dE)$.
Let us consider $g(E)$ of the general form
\be
g(E) = c_0 V E^\a,
\ee
where $c_0$ and $\a$ are constants. This covers the case of
non-relativistic particles with $\a =1/2$,
$c_0 = 2^{1/2} m^{3/2}/ \pi^2$
and the relativistic case with
$\a =2$, $c_0 = 1/\pi^2$, where the spin factor $g_s =2s+1 =2$ has been
included.
This gives the total number of states
$N_S = \int_0^\mu d E g(E) $ 
\be \label{N}
N_S = \frac{c_0 V}{\a+1} \m^{\a+1}
\ee
and the total internal
energy $U = \int_0^\mu dE
g(E)  E $
\be \label{U}
U =  \frac{c_0 V}{\a+2} \m^{\a+2},
\ee
where $\mu$ is the Fermi  energy.

As a result, the response of the energy to change in volume gives rise to an
``internal energy pressure'' $P_U := - (\del U /\del V)_N$ 
\be\label{Pm-mu}
P_U =  -  \frac{\a+1}{\a+2} \; N_S \left(\frac{\del \mu}{\del V}\right)_N.
\ee
In this system, the total number of fermions is given by $N= 2 N_S$
since each energy state is occupied by two fermions with opposite
spins. Therefore, the degeneracy pressure is determined by the volume
dependence of the Fermi level $\mu$. Note that from \eq{U}, $\mu$ is
approximately equal to the average energy of the states. Note also
that the form of $P_U$ depends crucially on the volume dependence of
the Fermi energy. We will see later that the Fermi energy \eq{ER} in
our black hole model has an entirely different behavior from that of a
system of ordinary fermionic particles, see, e.g.,
\eq{mu-NR_part}. This
crucial difference allows our system of Fermi degrees of
freedom to model a black hole.

\subsection{Matter degeneracy pressure: neutron star}

As a warm up, let us apply this model to consider the
example of a non-relativistic neutron star ($\a=1/2$). It follows from
\eq{N} that 
\be \label{mu-NR_part}
\mu = \frac{1}{2m}\left(\frac{3\pi^2 N}{V}\right)^{2/3}
\ee
and hence
\be \label{Pm}
P_U \sim \frac{ N^{5/3}\hbar^2}{m R^5},
\ee
where $\sim$ means equality up to a proportional constant of order 1.
Here, the total number of particles is given by
\be \label{Nm}
N= M/m
\ee
where $m$ is the mass of the neutron.
In the case of a neutron star, the internal energy and
the pressure $P_U$ arises from the kinetic motion of the neutrons.
Now for a neutron star of mass $M$ and radius $R$, the gravitational energy
is $E_g = - \g GM^2/R$ where $\g$ is a constant which depends on the
distribution of the masses. 
This gives the
gravitational pressure
$P_g = -\frac{\del E_g}{\del (V_0 - V)} = {\del E_g}/{\del V}$,
where $V= 4\pi R^3/3$ is the volume of the star 
and $V_0$ is the fixed volume of an infrared box where the star sits in.
We obtain the gravitational pressure
\be \label{Pg}
P_g = \frac{\g}{4\pi} \frac{G M^2}{R^4}
\ee
acting on the star. 
 The star settles at a radius when equilibrium $P_g = P_U$ is attained.
This gives immediately the mass-radius relation
\be
R \sim M^{-1/3}.
\ee
The requirement that $R$ is
greater than the Schwarzschild radius gives 
the mass limit $M \lesssim 2 M_{\odot} $ for neutron star.

One may wonder if one could achieve the black hole mass-radius relation \eq{RM}
with some 
choice of $\a$. The answer is no. In fact, as long as $N$ is given
by \eq{Nm} for a fixed mass $m$ of the fermions, 
the scaling relation is
\be
R \sim M^{- \a/(2-\a)}
\ee
and it clearly shows that no choice of $\a$ can reproduce the black hole
mass-radius relation. This is just a simple way to see that a Fermi system of
particles where \eq{Nm} holds can never model a black hole.
In addition to this,
the usual Fermi gas model as described above
also seems not to be able to account for the
Bekenstein-Hawking entropy. In fact, in the standard analysis, 
the degenerate Fermi gas is in a ground state obtained
by filling up all the energy levels up to the Fermi energy in
accordance with the Pauli exclusion principle. This gives 
a unique ground state and hence a vanishing microstate entropy
\be
S = \log \cG =0,
\ee
where $\cG = 1$ is the ground state degeneracy.
In the next subsection, we will see that both problems can be resolved
together by considering a system of Fermi 
quanta whose
Fermi energy
has a simple dependence on the size of the system.

\subsection{Schwarzschild black hole as a system of Fermi quanta}

The fact that a black hole has an entropy means that a black hole is
not a classical vacuum as described in general relativity but a
nontrivial quantum system of microstates. Obviously, these microstates
cannot arise from the ordinary elementary particles of the standard
model since, according to the singularity theorem, the ordinary matter
energy-momentum cannot withstand gravity collapse. Instead, the
universal nature of black holes suggests that these quanta should have
a generic identity independent of the matter that has been collapsed
to form the black hole. It then seems natural to consider the black
hole interior a region filled with elementary quanta of the quantized
spacetime. As a Fermi system is naturally equipped with a degeneracy pressure, which has the potential to counter the collapse and keep the system from further collapsing under the force of gravity, let us, therefore, consider a system of fermionic degrees of freedom in a spherical volume of radius $R$ as a model of the interior spacetime of black hole.
As we will show now, by taking $R$ to be the horizon radius of the black hole and with a specification of its density of eigenstates,
this simple Fermi system reproduces all the desired properties of the quantum black hole.

Outside the black hole, Einstein gravity and
the Schwarzschild metric is a good description of the long range physics.
From the outside observer point of view,
the Schwarzschild black hole is described by a vacuum solution of the
Einstein equation with a horizon radius $R=2GM$ where $M$ is the mass of the
black hole.  In general, given an asymptotically flat spacetime
with a timelike Killing vector $t^\a$, the 
energy of the solution is given by the Komar mass 
\be \label{komar-def}
M = - \frac{1}{8\pi G} \int_{S_\infty} \nabla^\a t^\b dS_{\a\b},
\ee
where $t^\a$ has been normalized such that $t^2 =-1$ asymptotically.
The energy $M$, however, contains both the gravitational energy and
other energy in the spacetime, e.g. rotational energy or Coulombic energy.
This motivates the adoption of a surface integral over the horizon
\cite{Poisson:2009pwt}
\be
E_g := - \frac{1}{8\pi G} \int_{H} \nabla^\a t^\b dS_{\a\b}
\ee
as the gravitational energy of the black hole. $E_g$ may be expressed in terms
of the surface gravity $\k$ of the horizon as
\be \label{Eg}
E_g := \frac{\k  A}{4 \pi G} 
\ee
For the Schwarzschild black hole, we have
\be \label{Eg-rR}
E_g = M.
\ee
This  give rises to the gravitational pressure
\be
P_g = \frac{\del E_g}{\del V} = \frac{1}{8 \pi G R^2}
\ee
which act on the Fermi system from outside. 
Here we have used $M = R/2G$ for the Schwarzschild black hole.
The Fermi system is stable if  the {\it gravitational pressure $P_g$
  from outside} can be balanced by some
{\it  internal pressure $P_U$ from of} of the black hole
\be \label{PP}
P_g = P_U.
\ee

To determine $P_U$, one needs to know the density of energy eigenstates $g(E)$
and the energy $\mu$ of the Fermi system. In the context of matrix model, the
fermionic energy eigenstates arises from a quantization of the
quantum mechanical
Hamiltonian in some background of the bosonic matrices which corresponds to
the black hole. In any case, it is natural that the energy eigenvalues are
related to the inverse size of the system.
Therefore let us propose a Fermi energy of the form
\be \label{ER}
\mu = \frac{a}{R},
\ee
where $a$ is some numerical constant to be fixed.
Substitute this into \eq{Pm-mu}, we obtain
the  pressure 
\be \label{PmR}
P_U = \frac{U}{4 \pi R^3}
\ee
due to the internal energy of the Fermi system. Note that it is independent
of $a$. 
Note also that $P_U = \frac{U}{3V}$ implies the Fermi system is Weyl
invariant, i.e.,
$\rho_U = 3P_U$.
Now if we assume the equivalence principle holds,
then the internal  energy $U$ of the Fermi system is equal to
the gravitational mass $M$ of the black hole 
\be \label{UM}
U=M.
\ee
From which we arrive immediately
at \eq{PP} and so our Fermi model for the interior
of the Schwarzschild black hole is stable
against the squash of  gravity.
This result is interesting and deserves a couple of remarks.
\ben
\item
In general relativity, the black hole is a vacuum solution with
vanishing energy density and vanishing pressure. With a vanishing
pressure, the squash by gravity is unavoidable. In our model,
the black hole is populated with
a set of fermionic quantum states and a nonvanishing degeneracy
pressure.

\item 
  The presence of a positive pressure is crucial to the balance
  of the system against gravitational collapse. A deeper
  understanding of the theory of
quantum gravity is needed to understand the nature of the quantum
spacetime of the black hole.

\item In our discussion above, we have assumed
  that the mass $M$ is sufficiently large so that the spacetime outside
  the black hole can be described semi-classically
  in terms of a metric. For smaller
  mass where the semi-classical approximation fails, the exterior region
  of the black hole should also be described quantum mechanically
  in terms of the Fermi quanta.

\item 
It is interesting to note that the Newton constant was not present in
the original expression \eq{PmR} but appears only in the
gravity description after the equivalence principle is invoked.
It suggests that gravity is, in fact,
an effective theory that emerges from an underlying fundamental theory
of quantum spacetime. The coupling constant for this fundamental
interaction of spacetime would be a dimensionless pure number.

\een

Next we look at the entropy.
We note from \eq{UM} that $U \propto R$.  This gives $\a=0$
in \eq{U} and hence
\be \label{Umu}
U = \frac{1}{2} N_S \mu
\ee
Or, on using \eq{UM} and \eq{RM}, we obtain the total number
of energy eigenstates 
\be \label{NRG}
N_S =\frac{1}{\pi a}\cdot \frac{\pi R^2}{G \hbar},
\ee
where we have restored $\hbar$  in order to emphasis the
quantum nature of \eq{NRG}.
We note that, interestingly,
\eq{NRG} has an area dependence and suggests that the
Bekenstein-Hawking entropy may find an explanation in our model. In fact
if somehow there is a power like ground state degeneracy
$\cG = ({\rm something})^{N_S}$, then the Bekenstein-Hawking entropy
would follow immediately
as a coarse grain entropy.

It looks however a little puzzling in our present picture since 
in the standard treatment of degenerate Fermi system, the 
ground state is solely dedicated by energetics and is obtained by 
filling up all of these $N_S$ states
by a pair of fermions with opposite spins 
$\ket{\pm}$.  This results in an unique ground state
wave function and zero entropy.
However this is not the only thing that can happen.
The fermion degrees of freedom may also be described additionally by
some bosonic wave functions. This is for example the case in an atom
where the electronic states are  given by a spin wave function
together with  an orbital wave function.
Therefore let us consider the simple setup where there are
two bosonic states $\chi_1, \chi_2$ corresponds to each
energy level. Then after taking into account of Fermi statistics,
there are four two-bodies wave functions 
\bea\label{wf}
& \chi_1(1) \chi_1(2)\times \ket{1,2}_-, \quad
\chi_2(1) \chi_2(2) \times \ket{1,2}_- ,\quad
\left(\chi_1(1) \chi_2(2) + \chi_1(2) \chi_2(1)
\right) \times \ket{1,2}_-, \nn \\
& \left(\chi_1(1) \chi_2(2) - \chi_1(2) \chi_2(1) \right) \times \ket{1,2}_+
\eea
where \be \label{ket-pm}
\ket{1,2}_\pm :=\frac{1}{\sqrt{2}} \big(\ket{+}_1 \ket{-}_2
\pm \ket{-}_1 \ket{+}_2 \big)
\ee
are the spin wave functions.
Let us 
denote the set \eq{wf} of wave functions by $\psi_{E, n_E}, n_E = 1,
\cdots, 4$.  Then  the ground state wave function is given by
\be \label{gs}
\Psi _{\{n_E\}} = \prod_{0<E<\mu} \psi_{E,n_E},
\ee
for a given specification $\{n_E\}$
of symmetry of the ground state.  We remark
that in the spinor space, the fermion
pairs at the same energy level are entangled in Bell states.
This explain the origin of the assumption made in  \cite{Chu:2022ieq}. 
We remark that since each energy level has a degeneracy of 4 as in \eq{wf}, 
the number of energy levels is given by
\be \label{NLNS}
N_L = N_S/4,
\ee
and since each state in \eq{wf} describes 2 fermionic degrees
of freedom, the number $N$ of fermionic degrees of freedom
contained in each of the $\cG$ ground states is given by
\footnote{Note that  we have the same \eq{NNL}
but $N_S = 2N_L$  for the neutron star system since each energy level
has a degeneracy of 2, e.g. as given by the basis \eq{ket-pm}.} 
\be \label{NNL}
N = 2 N_L
\ee
and
\be \label{NsNg}
N =   \frac{A}{4G} \times \frac{1}{2\pi a}.
\ee
This is in sharp contrast with
the expression \eq{Nm} for ordinary particles.
\begin{figure}
  \centering
      \includegraphics[width= 8cm]{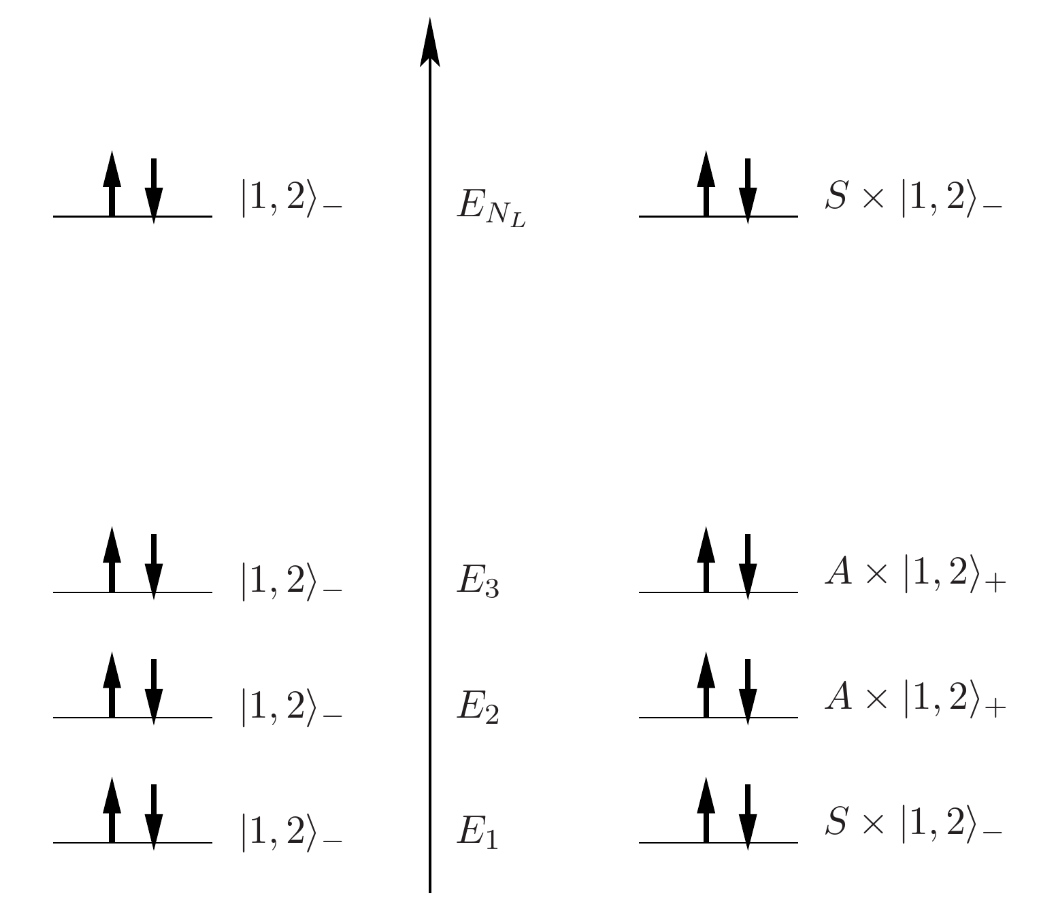}
      \caption{Left: Typical non-degenerate ground state of Fermi system.
        Right: An entangled ground state of spacetime of black hole.}
\end{figure}
Now as there is a choice
of 4 states
$\psi_{E,n_E}$ for each energy level $E$, the ground state \eq{gs}  has a
degeneracy of
\be
\cG = 4^{N_L} = 2^{N_S/2}
\ee
The coarse graining of these entangled microstates then give rises to the
entropy
\be
S = \log \cG = N_S/2,
\ee
where a base 2 logarithm is used.
This reproduces the Bekenstein-Hawking entropy if
$ a = \frac{1}{2 \pi }$.
As a consistency check, \eq{NRG} can also be  reproduced from \eq{N} if
$c_0 = 3\pi/G$.
Summarizing,  our model is defined by the Fermi level
\be \label{mu}
\m = \frac{1}{2\pi R}
\ee
and a constant density of energy eigenstates
\be \label{gE}
g(E) = \frac{3\pi V}{G}. 
\ee

We emphasis that the density of energy eigenstates $g(E)$ is not to be confused
with the density of microstates $\omega(E)$, which appears directly in
the counting of the coarse grain entropy.
In the above, we have derived
that a constant $g(E)$ give rises to the needed number of microstates
to account for the Bekenstein-Hawking entropy.
In the context of matrix model, e.g. BFSS, it is interesting to find out what 
background of the bosonic matrices would give rises to a fermionic quantum
mechanics with a constant density of eigenstates.

In general
any classical spacetime should admit a fermionic description at the fundamental
quantized level.
It is an interesting question to understand how a quantum
spacetime other than a quantum black hole is characterized by a density
of states $g(E)$. It is often assumed
that black hole is the closest packed object in nature. This would imply
an upper bound on $g(E)$ for an arbitrary quantized spacetime. Since the
maximal allowed energy is the Planck scale
\footnote{The precise value 
of the Planck scale cutoff is not essential,
and is chosen here for purely convenience
so that the bound \eq{Nmax} looks simpler.}
$\Lambda_P = 1/(3\pi l_P)$, we obtain an upper bound on the
number of energy eigenstates $N_S = \int dE g(E)$:
\be \label{Nmax}
N_S \leq  \frac{V}{l_P^3}
\ee
which is a statement on the {\it maximal capacity of states} in any volume $V$.
Here we have used $G=l_P^2$ where $l_P$ is the Planck length.
Due to the general character of
this statement, we  propose this to
be a fundamental principle of quantum gravity.
A possible origin of the upper bound \eq{Nmax} of states is that
there is a lower limit on the locality in space
\be \label{ur}
\D V \gtrsim l_P^3.
\ee
Due to \eq{ur}, quantum gravity is manifestly nonlocal below the Planck length
scale.
This provides a way out of the usual singularity theorem and
suggests that our model is in fact consistent without singularity.
It will be interesting to
understand deeper the meaning and origin of \eq{ur}. It may be
related to some kind of uncertainty relation of the quantized
spacetime.

We note that the above construction of the ground state wave function can
be easily generalized to the case when there is a bosonic degeneracy of $g$
for each energy level $E$.
Denote the single particle wave functions by $\chi_a$,
$a=1,\cdots, g$, then after taking into account of Fermi statistics,
the two bodies wave functions are given by
\be \label{wf-g}
S_{ab}(1,2)\times \ket{1,2}_-,
\qquad
A_{ab}(1,2)\times \ket{1,2}_+,
\ee
where
$S_{ab}$ (respectively $A_{ab}$) are the symmetric (respectively antisymmetric)
2-bodies wave functions formed by  $\chi_a, \chi_b$.
There are in total $g^2$ of them. The ground state wave function of the system
is given by \eq{gs} and there is a degeneracy of $\cG = (g^2)^{N_L}$ where the
number of energy level is now $N_L = N_S/g^2$.
This give rises to the entropy $\sim 2N_S/g^2$ and is equal to \eq{BH}
for $g=2$. 

We note also that our construction of entangled ground state wave
function can be generalized easily to ground state wave function with
more complicated entanglement, e.g. multi-bodies. In principle, it
could mean that black holes can exist in other quantum states with
different entropy content. But it may also mean that these more
general entangled states are actually forbidden in quantum
gravity. Formulating the selection rule and understanding how and why
it works would give useful hints on the properties of quantum gravity.

We remark that in our model, the area dependence of the
Bekenstein-Hawking entropy is explained by a set of fermionic degrees
of freedom, which are gravitational in origin and live in the interior of the
black hole. On
the other hand, in the picture of \cite{Susskind:1994vu}, the
explanation is found in terms of a two-state spin system associated
with each site of a two-dimensional lattice of quanta living on the
horizon area. The two descriptions are supposed to be related by
holography. In our model, the Bekenstein-Hawking entropy originates in
the entanglement of the fermionic degrees of freedom. This is nothing
but another illustration of the celebrated observation of Ryu and
Takayanagi \cite{Ryu:2006bv} that entanglement is at the heart of
holography.  Our model suggests that holography is a consequence of
the fermionic nature of the quantum spacetime.

We remark that the area law receives corrections in higher derivative modified
gravity. In principle, such corrections can be accounted for
in our model by using a modified Fermi level
and a modified density of states. However deriving these results will require
the knowledge of a quantum theory of gravity.
Please see \cite{Chu:2024qil,Chu:2024edh} for a recent proposal of a theory
of quantum gravity as a quantum mechanics of quantum spaces. It is
quite amusing that most of the
features of the phenomenological Fermi model appear there in a consistent
setup.

Finally, let us comment on the first law of black hole
thermodynamics. 
Using \eq{Umu}, \eq{NLNS}, \eq{NNL} in our model, we find $U= N \mu $ for
the energy of the Fermi system, where the number of
fermions $N =A/4G$ and $\mu =1/(2\pi R)$. This gives
immediately
\footnote{
We take $R$ and $N$ (i.e. $G$) as independent
variables in the derivation below.
That $G$ is allowed to  vary independently follows from
the point of view that the Newton constant
$G$ can be taken as the expectation value of some background field. A
similar trick has been utilized to argue for the holomorphic dependence of the
gauge coupling in supersymmetric gauge theories. 
}
the pressure
\be \label{P2}
P_U = - \left(\frac{\del U}{\del V}\right)_{N,S} = \frac{1}{8 \pi G R^2}  
\ee
and the chemical potential
\be \label{muC}
\mu_C = \left(\frac{\del U}{\del N}\right)_{V,S} = \mu.
\ee
As a consistency check, due to the fact that $S=N$ in our model,
we get a vanishing temperature
\be
T =  \left(\frac{\del U}{\del S}\right)_{V,N} =0,
\ee
which is in fact what we have considered for a completely degenerate system.
It is easy to
see that  the following relation holds
\be \label{1st-our}
dU = - P_U dV + \mu_C dN.
\ee
At this point, it might be a bit
puzzling as to why \eq{1st-our} is different from
the usual statement of black hole thermodynamics
\be \label{1st-hor}
dU = T_H dS,
\ee
which involves only the entropy term, and has neither a 
pressure term nor a chemical potential term as in \eq{1st-our}.
The puzzle is resolved once we realize that the
form \eq{1st-hor} of the first law was deduced from the physics
outside the black hole, e.g., from the properties
of the Hawking radiation or from the
Brown-York stress tensor defined  outside
the black hole \cite{Brown:1992br}.
In particular, $T_H$ is the temperature of the Hawking radiation
and \eq{1st-hor} is the first law of the black hole as observed by
an  asymptotic 
observer based on the degrees of freedom {\it outside} the black hole.  In
comparison,  \eq{1st-our} is the first law for the {\it interior} of the
black hole. The existence of a nonvanishing pressure \eq{P2} and chemical
potential \eq{muC} follows from the properties of the fermionic
degrees of freedom of the black hole. Note that while classically not
even light can escape a black
hole, it is possible to access the interior of a black hole through
quantum mechanical effects such as tunneling or entanglement.  For
example, one can create an entangled pair of states and allows one of
them to fall into the black hole. With careful handling of the state,
the entanglement can be preserved. This would allow the outside world
to probe the interior of the black hole through entanglement
measurement.

\subsection{Reissner-Nordstr\"om charged black hole}

Next, let us consider a charged black hole with mass $M$,
charge $Q$. In general relativity, it is described by
the Reissner-Nordstr\"om metric with the horizon radius $R$ satisfying
the quadratic relation
\be \label{RRR}
R^2 - R R_S + R_Q^2 =0,
\ee
where $R_S = 2GM$ and $R_Q = \sqrt{GQ^2}$. We now show that
the properties of the charged black hole can also be reproduced
by the Fermi system. 
From our analysis in the last subsection, we have learned that
the system of Fermi quanta has a constant density of states
\be \label{p1}
g(E) = \frac{3 \pi V}{G}
\ee
and a Fermi energy
\be \label{p2}
\mu = \frac{1}{2\pi R}.
\ee
It is natural to assume that these properties are fundamental
quantum properties of black hole spacetime and holds
also here.
As a result, we obtain the total number of states
\be \label{NGRq}
N_S = \frac{2\pi R^2}{G},
\ee
the internal energy of the Fermi sea
\be \label{Uq}
U = \frac{1}{2} N_S \mu = \frac{R}{2G}
\ee
and the internal energy pressure $P_U = - (\del U/\del V)_N$,
\be \label{Pmq}
P_U = \frac{1}{8 \pi G R^2}.
\ee
We note that \eq{NGRq}, \eq{Uq} and \eq{Pmq} are obtained 
from the properties \eq{p1}, \eq{p2} and are independent of the
charge.

Next for the gravitational pressure. Using \eq{Eg},
the gravitational energy of the charged black hole is given by 
\be
E_g = \frac{R}{G} -M.
\ee
In addition, 
there is also a Coulombic
energy $U_Q$ due to the presence of an electric field. We can compute
$U_Q = \int_R^\infty u \; 4\pi r^2 dr$
from the electric
energy density $u = E^2/(8\pi) = Q^2/(8 \pi r^4)$ outside the black hole,
this gives
\be
E_Q = \frac{Q^2}{2R}.
\ee
Invoking the equivalence principle, the total energy $E_g+E_Q$
of the black hole solution should be equal to the internal energy $U$ of
the Fermi system.
This gives precisely the mass-charge-radius relation
\eq{RRR}, which in turn implies that
\be
E_g = \frac{R}{2G} - \frac{Q^2}{2R}.
\ee
This leads to the
gravitational pressure
\be
P_g = \frac{1}{8 \pi G R^2} + \frac{Q^2}{8 \pi R^4}
\ee
acting from outside on the horizon of the charged black hole.
Added to it is the  electromagnetic pressure $P_Q = \del E_Q/\del V$
\be
P_Q = - \frac{Q^2}{8 \pi R^4}.
\ee
As a result we find a balance of pressure $P_U = P_g+P_Q$
for the Fermi system
and our model for the black hole 
is stable against the gravitational squash.

We remark that the radius relation \eq{RRR} was originally derived from the
condition $1/g_{rr}|_{r=R} =0$ at the horizon of the
Reissner-Nordstr\"om metric
\be
ds^2 = - \Big(1-\frac{R_S}{r} + \frac{R_Q^2}{r^2}\Big) dt^2 +
\Big(1-\frac{R_S}{r} + \frac{R_Q^2}{r^2}\Big)^{-1} dr^2 + r^2 d\Omega^2.
\ee
As such, there is no simple understanding of \eq{RRR} apart
from the fact that it comes from the Einstein equation. 
It is quite remarkable that in our model of the black hole,
the radius relation \eq{RRR} admits a direct physical 
interpretations in terms of the equivalence principle, which in
turns guarantee  the stability of the black hole model.
We also note that the counting for the Bekenstein-Hawking entropy works the
same way as discussed above. 

\subsection{Black hole with a cosmological constant}

In this subsection, we show that our model also works for black holes with
a cosmological constant. Let us start first with the de Sitter space
\cite{Spradlin:2001pw}
with a positive
cosmological constant $\L >0$. In the static coordinates, the metric
\be \label{dS}
ds^2 = - (1-\frac{\L r^2}{3}) dt^2 +
(1-\frac{\L r^2}{3})^{-1} dr^2 + r^2 d\Omega^2, 
\ee
is a solution to the Einstein equation with a vacuum background energy
density $\r_\L = \L/(8 \pi G)$ and a background pressure
\be \label{bkgd-P}
P_\L =  - \frac{\L}{8 \pi G}.
\ee
%
The de Sitter black hole is a solution in this background with the metric
\be \label{dS-BH}
ds^2 = - (1-\frac{2GM}{r} - \frac{\L r^2}{3}) dt^2 +
(1 -\frac{2GM}{r} -\frac{\L r^2}{3})^{-1} dr^2 + r^2 d\Omega^2.
\ee
For $M <M_c$ less than a certain critical mass $M_c$, the condition
\be \label{dSBH-R}
1-\frac{2GM}{r} - \frac{\L r^2}{3} =0
\ee
has two positive roots, giving the location of the black hole horizon
and
the cosmological horizon, respectively. At $M =M_c$, the two
positive roots meet, meaning that there is a
maximum size black hole
which can fit of the de Sitter space before the black hole horizon
hits the
cosmological horizon.

Let us consider the case $M<M_c$ of interest. Let $R$ be the black
hole horizon radius and $L$ be the cosmological horizon radius.
The black hole interior region $0<r<R$ is described by a system of quanta
with the Fermi energy and the density of states
\be \label{fR}
\m = \frac{1}{2\pi R}, \qquad g (E) = \frac{3 \pi}{G} V_R,
\ee
where $V_R := 4\pi R^3/3$.
This gives the internal energy
\be \label{fR1}
U = \frac{R}{2G}
\ee
for the Fermi system and the degeneracy pressure
\be \label{fR2}
P_U = \frac{1}{8 \pi G R^2}.
\ee
In addition, the system has an energy
\be
E_\L = \frac{\L}{6G} R^3
\ee
and the pressure \eq{bkgd-P} due to the presence of a cosmological constant.
As for the gravitational energy,  we note that the definition  \eq{Eg}
requires an asymptotic flat background as a reference level for the energy.
Now since  there is no asymptotically flat region in the de Sitter black
hole spacetime
and also because there is not a timelike Killing vector in the global de Sitter
spacetime, the definition of mass in de Sitter spacetime is non-trivial.
One can employ a convenient definition of the black hole
mass by  subtracting away  the background contribution from the pure
de Sitter space ($M=0$): 
\be \label{subt}
E_g := \frac{\k A}{4 \pi G} - \left(\frac{\k A}{4 \pi G}\right)_{M=0}.
\ee
This gives for the dS black hole
\be \label{EgM}
E_g = M.
\ee
Actually one can use any other definition as long as the
relation \eq{EgM} is reproduced.
The principle of equivalence then states that equality $U = E_g + E_\L$
between the energy of the Fermi system and the total gravitational energy
of the black hole solution. This
gives immediately the relation \eq{dS-BH} satisfied by the horizon radius
of a dS black hole.
At the same time, we have from \eq{EgM} the gravitational pressure
\be
P_g = \frac{1}{8 \pi G R^2} -\frac{\L}{8 \pi G}
\ee
in the outside region of the black hole.
Note  that the background pressure of de Sitter space is automatically
reproduced here. All in all,
the pressure is balanced:
\be
P_U = P_g + (-P_\L),
\ee
where the term $-P_\L$ is the pressure acting on the system from externally
due to the background energy density.
Our model for
the black hole is stable.
We remark that as the Fermi degrees of freedom is supposed to be elementary
quanta of the quantized space, it should be possible to build a Fermi model
that describes the cosmological horizon as well. This is an interesting problem
to consider for the quantum mechanics of quantum space
\cite{Chu:2024qil,Chu:2024edh}.

Finally, let us consider the case of the AdS black hole. Classically,
the metric is given by
\be \label{AdS-BH}
ds^2 = - (1-\frac{2GM}{r} +\frac{|\L|\, r^2}{3}) dt^2 +
(1 -\frac{2GM}{r} +\frac{|\L| \, r^2}{3})^{-1} dr^2 + r^2 d\Omega^2.
\ee
In this case, the black hole horizon radius satisfies the relation
\be \label{AdSBH-R}
1- \frac{2GM}{R} + \frac{|\L| R^2}{3} =0
\ee
and the black hole is situated in the background AdS space
with a boundary at infinity.
We model the black hole interior with a Fermi system of the same properties
\eq{fR}, \eq{fR1}, \eq{fR2}. The 
gravitational energy for AdS black hole is
well defined and one have
\be
E_g =M.
\ee
Equating the energy of the Fermi system
$\frac{R}{2G}$ with the total gravitational energy $M - \frac{|\L|}{6G} R^3$
of the black hole spacetime, we obtain the mass-horizon radius relation for AdS
black hole. In the same way, we have the gravitational pressure 
\be
P_g =  \frac{1}{8 \pi G R^2} +\frac{|\L|}{8 \pi G}
\ee
in the outside region of the black hole.
Pressure is again balanced and the Fermi model of black hole interior is stable
against collapse. The Bekenstein-Hawking entropy works out the same way
as before.

\section{Conclusion and Discussions}

In this paper, we have proposed a quantum mechanical model of the black
hole as a system of fermionic degrees of freedom residing within the
horizon. Quite amazingly, with only a few simple assumptions, our
bottom-up approach can resolve immediately all the mentioned puzzling
properties of black holes, all at once in the same model. In our
model, there is a pressure of the black hole, which balances out
the gravitational collapse, and there is a chemical potential that
measures the amount of energy carried away by the Hawking radiation
quanta. The temperature is zero, yet a Bekenstein-Hawking entropy
arises due to the presence of entanglement of states. We also find
a new expression of the black hole mechanics in terms of the black
hole interior degrees of freedom. We
show that the fermionic nature of the black hole ground state,
together with its entanglement structure is crucial in 
obtaining of a nonsingular interior and providing the needed microstates
counting.


In general, we propose that quantum gravity can be described in terms
of a fermionic quantum mechanics.  In this paper, we took a
bottom-up approach to build the quantum mechanics
by invoking only what is necessary
\footnote{For example,  supersymmetry which is built-in
in the BFSS matrix model plays no role here.}.
According to this
picture, ordinary spacetime and gravity should arise as some effective
description and interaction of the Fermi variables, with the Newton
constant emerges in the gravitational description.  Accordingly, a
general spacetime would have an entangled wave function associated
with it. It is interesting to ask whether and how the entanglement
asset contained in spacetime is observable in general.  It is also
interesting to understand how the Einstein equation may emerge as
effective dynamics of the underlying fermionic degrees of freedom,
perhaps due to the entropic property of these variables.

We comment that the Page curve behavior of the entanglement entropy of
Hawking radiation may be analyzed in our model. In \cite{Chu:2022ieq},
the set of Bell states was assumed to be located in a thin layer
underneath the horizon, and it was shown that tunneling results in the
Page curve. In our present model, the entangled states are distributed
all over the
``interior''. Nevertheless, the same analysis can be carried
out, and one can show that the tunneling process also resulted in the
Page curve \cite{Chu:2023agv}.

We remark that despite the fact that our model is stable against
gravitational collapse in
terms of pressure, it is necessary to
understand more precisely how the prediction of singularity in general
relativity is avoided in our model.
This may be related to our proposal that the interior of black hole
is a quantum spacetime with a
nonvanishing lower bound \eq{ur} on the spatial nonlocality.
In fact, the relation  \eq{ur} suggests that spacetime at the
Planck scale is characterized by some form of noncommutative geometry.
A better understanding of this fundamental issue  is crucial but is
beyond the scope of the current work.

As matter collapses and forms the black hole, the energy
and information of the collapsed matter are supposed to be transformed
into the energy and entanglement of the quantized spacetime
stored and encoded by the fermionic variables. Recently, based on
observations on the growth rate of astronomical black holes in the
universe, the authors of \cite{Farrah:2023aaa,Farrah:2023opk} proposed
that dark energy need not be a property of the vacuum, but a black
hole could be a source of dark energy.
It is interesting to study further the properties of this
kind of dark energy using our model. It is also interesting to study
the response to gravitational wave \cite{Oshita:2022pkc}.

We remark that our proposal shared some similarities with the fuzzy
ball proposal \cite{Mathur:2005zp} in that the classical vacuum of
the black hole is replaced by a set of quantum states: fuzzy ball of
strings in the fuzzy ball proposal and the set of entangled fermionic
states in our proposal. Our proposal is more generic in character. It
will be interesting to understand better the differences and possible
connections. It will also be interesting to understand how our model
may arise from existing proposals of quantum gravity, such as string
theory or matrix model. In particular, a fermionic string theory may
provide the right framework in which the quantum geometry of the
black hole and the considered fermionic degrees of freedom may be
obtained as string quanta from the string quantization.
\cite{Chu:1998qz,Chu:1999gi, Schomerus:1999ug, Chu:1999ne,
  Blumenhagen:2011ph}.

Finally, we emphasize that our model of the black hole is just a first
step toward the theory of quantum spacetime. Historically, the Bohr
atomic model, with its postulate about the existence of quantized
orbits, has given valuable hints to the development of quantum
mechanics. It is hoped that
our simple model of black holes
did have provided
useful clues to the construction of the theory of
quantum gravity.

\section*{Acknowledgments}
C.S.C like to thank Nick Dorey,
Dimitris Giataganas, Pei-Ming Ho, Hikaru Kawai and  Himanshu Parihar
for helpful discussion. 
C.S.C acknowledge support of this work by NCTS and
the grant 110-2112-M-007-015-MY3 of the National 
Science and Technology Council of Taiwan. 
R.X.M thank support by National Natural Science Foundation of China
(No.12275366).

\bibliography{references}

\end{document}